\documentclass{article}

\usepackage{PRIMEarxiv}

\usepackage[utf8]{inputenc} % allow utf-8 input
\usepackage[T1]{fontenc}    % use 8-bit T1 fonts
\usepackage{hyperref}       % hyperlinks
\usepackage{url}            % simple URL typesetting
\usepackage{booktabs}       % professional-quality tables
\usepackage{amsfonts}       % blackboard math symbols
\usepackage{nicefrac}       % compact symbols for 1/2, etc.
\usepackage{microtype}      % microtypography
\usepackage{lipsum}
\usepackage{fancyhdr}       % header
\usepackage{graphicx}       % graphics
\usepackage{amssymb}
\usepackage{amsmath}
\usepackage{float}
\usepackage{tabularx}
\usepackage{booktabs}
\usepackage{longtable}
\usepackage{threeparttable}

\graphicspath{{media/}}     % organize your images and other figures under media/ folder

\newcommand{\orcid}[1]{\href{https://orcid.org/#1}{\includegraphics[scale=0.06]{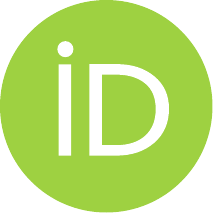}}}

\title{Bridging Equilibrium and Kinetics Prediction with a Data-Weighted Neural Network Model of Methane Steam Reforming
}

\author{
Zofia Pizoń\orcid{0009-0004-8964-0625} \\
AGH University of Krakow \\
Krakow, Poland \\
\texttt{zofiapizon@student.agh.edu.pl} \\
\And
Shinji Kimijima \\
Shibaura Institute of Technology \\
Tokyo, Japan \\
\texttt{kimi@shibaura-it.ac.jp}
\And
Grzegorz Brus\orcid{0000-0003-4911-5880} \\
AGH University of Krakow \\
Krakow, Poland \\
\texttt{brus@agh.edu.pl} \\
}

\begin{document}
\maketitle

\begin{abstract}
Hydrogen's role is growing as an energy carrier, increasing the need for efficient production, with methane steam reforming being the most widely used technique. This process is crucial for applications like fuel cells, where hydrogen is converted into electricity, pushing for reactor miniaturization and optimized process control through numerical simulations. Existing models typically address either kinetic or equilibrium regimes, limiting their applicability. Here we show a surrogate model capable of unifying both regimes. An artificial neural network trained on a comprehensive dataset that includes experimental data from kinetic and equilibrium experiments, interpolated data, and theoretical data derived from theoretical models for each regime. Data augmentation and assigning appropriate weights to each data type enhanced training. After evaluating Bayesian Optimization and Random Sampling, the optimal model demonstrated high predictive accuracy for the composition of the post-reaction mixture under varying operating parameters, indicated by a mean squared error of 0.000498 and strong Pearson correlation coefficients of 0.927. The network's ability to provide continuous derivatives of its predictions makes it particularly useful for process modeling and optimization. The results confirm the surrogate model's robustness for simulating methane steam reforming in both kinetic and equilibrium regimes, making it a valuable tool for design and process optimization.
\end{abstract}

% keywords can be removed

\section{Introduction}
\label{intro}

The significance of hydrogen as an energy source continues to grow along with its production. According to \textit{Global Hydrogen Review 2024} \cite{iea}, global hydrogen production reached 97 Mt in 2023, marking a 2.5 \% increase compared to 2022. The increasing demand for hydrogen comes from its versatility as an energy carrier, with the ability to be efficiently converted into electricity, making it a valuable medium for energy transport and storage \cite{rosen}. Hydrogen can be produced using both renewable and non-renewable energy sources. Currently, most of hydrogen \textendash ~over 60 \% \textendash ~is produced from natural gas without carbon capture, utilization and storage (CCUS) \cite{iea}, primarily due to its lower cost.

The most widely used method for hydrogen production from natural gas is methane steam reforming (MSR). This well-established process, first described in 1932 \cite{charles}. It does not require an external oxygen supply and operates at relatively lower temperatures compared to other hydrogen production methods. Furthermore, the post-reaction mixture consistently yields a hydrogen-to-carbon monoxide ratio greater than one, which is advantageous for downstream applications. However, MSR is associated with significant greenhouse gas emissions, making it one of the most carbon-intensive hydrogen production techniques \cite{HOLLADAY2009244}. Additionally, the MSR process requires a substantial energy input due to its endothermic nature. It also necessitates an excess of steam to prevent carbon deposition on the catalyst, which could otherwise hinder the reaction efficiency and reduce the longevity of the catalytic material \cite{en13040813}. The process can be represented by primary reactions: the Methane Steam Reforming Reaction (MSRR) the Water-Gas Shift Reaction (WGSR) and the Direct Reforming Reaction (DRR):

\begin{align}
    \mathrm{CH_4+H_2O} &= \mathrm{3H_2+CO},
   \label{MSR}
\end{align}
\begin{align}
     \mathrm{CO+H_2O} &= \mathrm{H_2+CO_2},
   \label{WGSR} 
\end{align}
\begin{align}
     \mathrm{CH_4+2H_2O} &= \mathrm{4H_2+CO_2}.
   \label{WGSR} 
\end{align}

One of the key applications of hydrogen is its use as a fuel in fuel cells, where it is converted into electricity. This application necessitates the miniaturization of reformers \cite{BAEK20149180}, driving a significant demand for process optimization. A promising approach to achieve such optimization involves leveraging methods based on artificial intelligence (AI), which can enhance efficiency and improve system performance. Artificial neural networks (ANNs) have been extensively utilized in modeling and optimizing hydrogen production processes, demonstrating their versatility and accuracy across various methods and operating conditions. Bilgiç et al. \cite{BILGIC202318947} highlighted the effectiveness of ANN-based approaches in hydrogen production modeling, showcasing their potential for improving process efficiency and prediction accuracy. Zamaniyan et al. \cite{ZAMANIYAN20136289} employed a backpropagation feed-forward ANN to model an industrial hydrogen plant, predicting temperature, pressure, and the mole fractions of hydrogen and carbon monoxide in the product stream. The input parameters included feed temperature, reformer pressure, steam-to-carbon ratio, and carbon dioxide-to-methane ratio in the feed. Similarly, Nasr et al. \cite{NASR20133189} applied a backpropagation ANN to predict hydrogen production from fermentation processes, using initial pH, substrate and biomass concentrations, temperature, and reaction time as inputs. Their model achieved a high correlation coefficient of 0.976.  Ghasemzadeh et al. \cite{GHASEMZADEH2018114} also developed a feed-forward ANN to compare the performance of silica membrane reactors and traditional reactors during methanol steam reforming. Their findings revealed that silica membrane reactors outperformed traditional ones in terms of methanol conversion, hydrogen recovery, and CO selectivity, with reaction temperature identified as the most influential parameter. In biomass gasification, George et al. \cite{GEORGE20189558} developed an ANN to model hydrogen production via air gasification. Their model, which predicted the composition of the post-reaction mixture based on seven operating parameters, demonstrated a regression coefficient of 0.987. For methane dry reforming over Ni/CaFe$_2$O$_4$ catalysts. Hossain et al. \cite{HOSSAIN201611119} compared multi-layer perceptron (MLP) and radial basis function networks. Statistical analysis indicated the superiority of the MLP network, with R$^2$ values of 0.9726, 0.8597, 0.9638, and 0.9394 for H$_2$ yield, CO yield, CH$_4$ conversion, and CO conversion, respectively. Similarly, Ayodele et al. \cite{https://doi.org/10.1002/er.6483} evaluated these two ANN architectures for hydrogen production from co-gasification of rubber and plastic wastes, finding that the single-layer MLP, with an R$^2$ of 0.990, provided the most accurate predictions. To further optimize hydrogen production, Ayodele et al. \cite{catal9090738} explored multiple ANN training algorithms, including the Levenberg–Marquardt algorithm, Bayesian regularization, and scaled conjugate gradient. The Bayesian regularization algorithm achieved the most accurate predictions, with the lowest standard error of estimates for CO and H$_2$ production rates in methane dry reforming. In subsequent work, Ayodele et al. \cite{AYODELE2021315} compared MLP-ANN and nonlinear response surface methods to estimate hydrogen concentration in reforming mixtures. The MLP-ANN model, trained on experimental data, outperformed the response surface method, achieving an R$^2$ value of 0.988. 

Building on these advances, other researchers have incorporated additional machine learning approaches, such as random forests and hybrid models, to further optimize process conditions. Nkulikiyinka et al. \cite{NKULIKIYINKA2020100037} employed both ANN and random forest algorithms to predict gas concentrations in reformer and regenerator reactors during sorption-enhanced steam methane reforming (SESMR). Their models, which utilized input parameters such as temperature, pressure, steam-to-carbon ratio, and sorbent-to-carbon ratio, achieved R$^2$ values exceeding 0.98, underscoring the accuracy of AI-driven approaches in hydrogen production modeling. Furthermore, Vo et al. \cite{VO2022820} optimized the integrated SESMR process using an ANN, combined with an economic model. Their findings revealed that hydrogen production at 99.99~\% purity could be achieved at a cost of 1.7 $\mathrm{\frac{\$}{kg}}$, 15 \% lower than production without carbon dioxide capture. ANNs have also been combined with advanced reactor modeling approaches. Dat Vo et al. \cite{VO2019113809} integrated an ANN with a multiscale model of the reactor, wall, and furnace used in the methane steam reforming process. The hybrid model was validated with reference data, achieving an error below 4 \%. Deterministic and stochastic simulations generated data for ANN training, enabling the creation of a rapid and accurate predictive tool for MSR operating parameters. These studies collectively underscore the versatility and effectiveness of ANNs in hydrogen production modeling and optimization, demonstrating their applicability across diverse production methods and reactor configurations.

This study builds upon the work of Pizoń et al. \cite{pizon}, where an ANN was successfully employed to predict the composition of the post-reaction mixture based on the operating parameters of the MSR process conducted under kinetic conditions. The objective of the present research was to extend the applicability of the model to include the equilibrium region, broadening its scope and utility.

\section{Methodology}
\label{methodology}
The development of an effective artificial neural network requires careful consideration of several critical factors. One of the most important aspects is the selection of an appropriate network structure. While complex architectures with numerous parameters can minimize training errors, they are prone to overfitting, where the model performs well on training data but poorly on unseen datasets. Another crucial element is the training process itself, which involves modifying the weights and biases of the network to accurately represent the given training data. It must strike a balance between duration and effectiveness. Training must be sufficiently prolonged to allow the network to learn, but excessive training (determined by the number of epochs) can lead to overfitting.

This study aims to design an appropriate ANN architecture to effectively simulate the methane steam reforming reaction. A well-prepared training dataset is equally essential for achieving reliable predictions. Accordingly, the initial phase of this research focuses on the systematic preparation of the dataset and the selection of an optimal network configuration to ensure accurate and robust modeling.

\subsection{Experimental setup}
\begin{figure}
    \centering
    \includegraphics[width=1\linewidth]{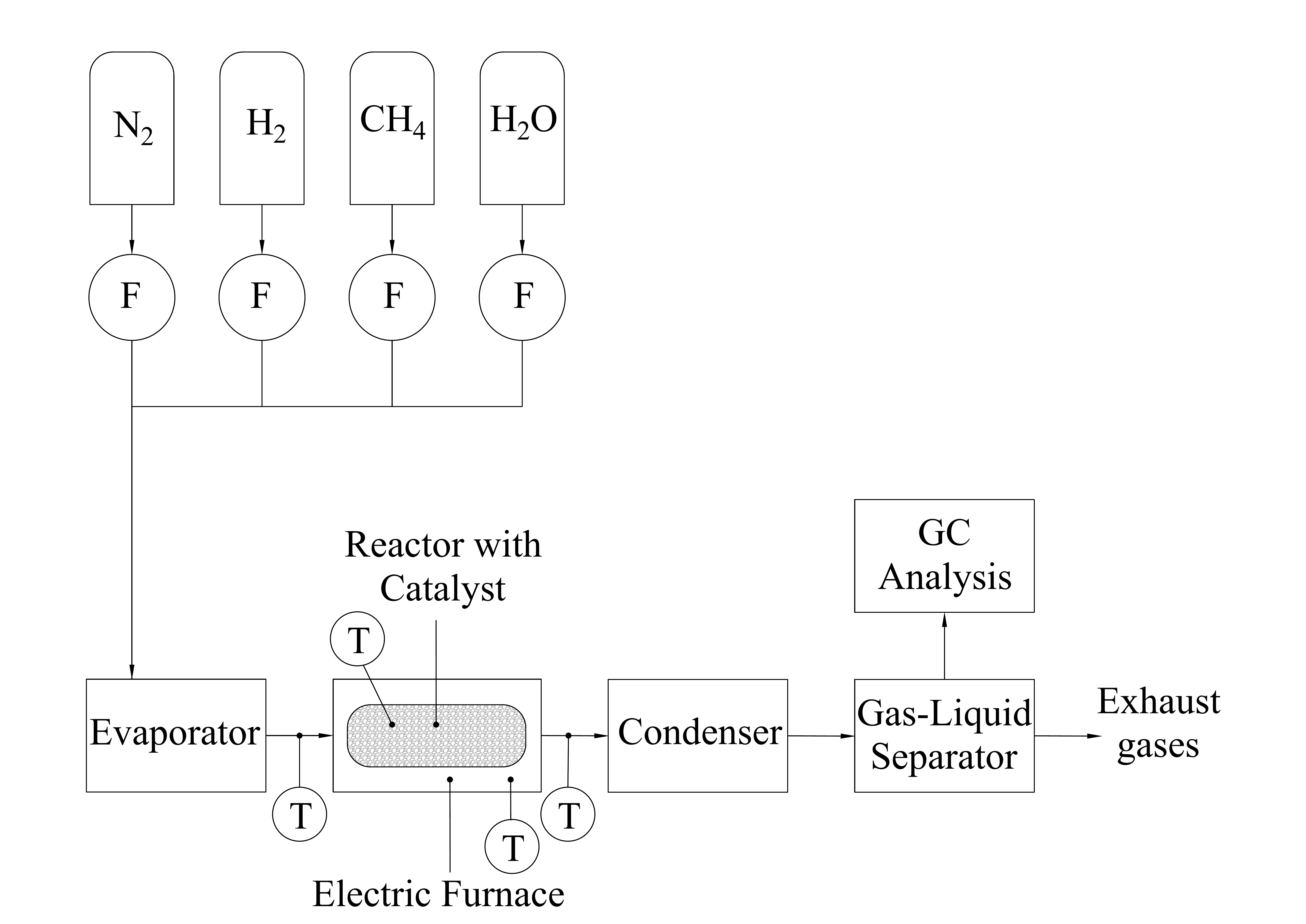}
    \caption{A scheme of experimental setup (F \textendash ~flow-meter, T \textendash ~thermocouple)}
    \label{exp_setup}
\end{figure}

The experimental data employed in this research were sourced from previous studies \cite{Brus.equil, brus.2012,Brus.2012b}. The experimental setup is illustrated in Figure \ref{exp_setup}. The core component of the setup was a stainless steel reformer, with geometric dimensions of 1 mm bed height, 25.4 mm radius, and 450 mm length, positioned within an electric furnace capable of reaching temperatures up to 1000~$^o$C. High-purity methane served as the fuel and was delivered to the reformer via a flow controller (denoted as F in Figure \ref{exp_setup}) and an evaporator that also functioned as a pre-heater. Water was supplied to the system using a pump. The reforming reaction tube was packed with a nickel catalyst supported on yttria-stabilized zirconia (NiO/YSZ), provided as a commercial product by AGC SEIMI CHEMICAL CO \cite{AGC2009Catalyst}. The catalyst, in the form of a fine powder, consisted of 60\% vol NiO and 60\% vol YSZ, with spherical particles averaging a diameter of 0.85 $\mu$m and a specific surface area of 5.2 $\rm{\frac{m^2}{g}}$. To activate the catalyst, a reduction process was performed before the experiment by heating it to 800 $^o$C for 6 hours in a mixture of 150 $\frac{\rm{ml}}{\rm{min}}$ nitrogen and 100 $\frac{\rm{ml}}{\rm{min}}$ hydrogen. To ensure a uniform thermal profile and minimize temperature gradients within the reformer, modifications to the reaction tube were implemented. The inlet zone of the reformer was partially filled with Al$_2$O$_3$ to preheat the incoming gas mixture to the desired reaction temperature. The thermal conditions of the setup were closely monitored using four thermocouples (denoted as T in Figure \ref{exp_setup}). The process was conducted at atmospheric pressure, and nitrogen was introduced as a diluent to reduce the methane conversion rate and ensure that the reaction occurs uniformly throughout the catalyst bed. Maintaining low methane conversion is crucial for obtaining an accurate reaction state. The obtained gas mixture was cooled to 2 $^o$C to condense in condenser and remove steam using gas-liquid separator. Then, it was analyzed through gas chromatography.

\subsection{Mathematical models}
In this study two mathematical models were used. The first one \cite{brus.2012, Brus.2012b} was a representation of methane steam reforming process in kinetic state. It was based on reaction rate equation: 
\begin{align}
    r_{\mathrm{st}}=w'_{\mathrm{cat}}\cdot k_{\mathrm{st}}\cdot p^a_{\mathrm{CH_4}} \cdot p^b_{\mathrm{H_2O}},
    \label{Rst}
\end{align}
where 
$r_{\mathrm{st}}$ \textendash ~reaction rate [$\mathrm{\frac{mol}{s\cdot m^3}}$], 
$w'_{\mathrm{cat}}$ \textendash ~catalyst density [$\mathrm{\frac{g}{m^3}}$], $p_{\mathrm{CH_4}}$ \textendash ~partial pressure of methane [Pa], 
$p_{\mathrm{H_2O}}$ \textendash ~partial pressure of steam [Pa],
$a=1.0$, $b=0.0$ \textendash ~dimensionless coefficients responding to the reaction order and $ k_{\mathrm{st}}$~\textendash ~reaction rate constant, wchich can be written using the following formula:
\begin{align}
    k_{\mathrm{st}}= A\cdot \exp{\left({-\frac{E}{\mathrm{R}T}} \right)},
    \label{kst}
\end{align}
where
$A=2.582 \cdot 10^{-4}$ $\mathrm{\frac{mol}{g\cdot Pa^{a+b}\cdot s}}$ \textendash ~pre-expotential constant,
$E=115255\ \mathrm{\frac{J}{mol}}$ \textendash ~activation energy,
$\mathrm{R} = 8.314472\ \mathrm{\frac{J}{mol\cdot K}}$ \textendash ~universal gas constant,
$T$ \textendash ~temperature of reaction [K],
The coefficients \begin{it} A, E, a, b \end{it} were estimated using a genetic algorithm tailored to match the experimental data. The equations to solve are derived from the conversion rate of the MSR reaction and the equilibrium equation for the shift reaction. The resulting pair of equations is shown in the following formulas:
\begin{align}
\label{eq:sys}
\left\{
\begin{array}{l}
x_{\mathrm{st}}=1- \dfrac{\left( \dot{n}^{\rm{inlet}}_{\rm{CH_4}}- r_{\rm{st}} \right) }{ {\dot{n}^{\rm{inlet}}_{\rm{CH_4}}} } \\
y \cdot \left( 3x_{\mathrm{st}}+x_{\mathrm{sh}}\right) -K_{\rm{sh}} \cdot \left(x_{\mathrm{st}}-x_{\mathrm{sh}} \right) \cdot \left( SC-x_{\mathrm{st}}-x_{\mathrm{sh}} \right)= 0
\end{array}
\right.
\end{align}
where: 
\begin{equation} 
\label{eq:K_sh}
K_{\rm{sh}}=\exp \left( -\frac{\Delta G_{\rm{sh}}^0}{RT} \right)
 \end{equation}
where $G_{\rm{sh}}$ \textendash ~the change of standard Gibbs free energy of shift reaction [$\mathrm{\frac{J}{mol}}$]. The equilibrium constant in Eq. (\ref{eq:K_sh}) is utilized within Eq. (\ref{eq:sys}) to determine the conversion rate of the shift reaction. The rate at which each chem (Eq. (\ref{WGSR})) is calculated based on the stoichiometry of the reactions, as detailed in Table \ref{tab:mole_change}.
\begin{table}[h]
\caption{Changes of chemical components inside the fuel reformer.\label{tab:mole_change}}
\begin{threeparttable}
\begin{tabular}{cccccc}
\hline
gas & $\mathrm{H_{2}O}$ & $\mathrm{CH_4}$ & $\mathrm{H_2}$ & $\mathrm{CO_2}$ & $\mathrm{CO}$       \\ \hline
inlet\tnote{a} & $SC$ & 1 & 0 & CC & 0  \\
steam ref. & $-x_{\mathrm{st}}$  & $x_{\mathrm{st}}$ & $-3x_{\mathrm{st}}$ & 0 & $x_{\mathrm{st}}$  \\
shift & $-x_{\mathrm{sh}}$  & 0 & $x_{\mathrm{sh}}$  & $x_{\mathrm{sh}}$  &               
$-x_{\mathrm{sh}}$                   \\
outlet & $SC-x_{\mathrm{st}}-x_{\mathrm{sh}}$ & $1 - x_{\mathrm{st}}$  & $3x_{\mathrm{st}}+x_{\mathrm{sh}}$ &$CC+x_{\mathrm{sh}}$ & $x_{\mathrm{st}}-x_{\mathrm{sh}}$  \\ \hline
\end{tabular}
\begin{tablenotes}
\item[a] For 1 mol of methane.
\end{tablenotes}
\end{threeparttable}
\end{table}

The second mathematical model \cite{Brus.equil} was a representation of methane steam reforming process in equilibrium state. It was based on equilibrium reaction constant equations presented in the following formulas:

\begin{equation} 
\label{eq:fgh}
\begin{cases} K_{\rm{st}} p_{\rm{CH_4}}  p_{\rm{H_2O}}  - p_{\rm{CO}} p_{\rm{H_2}}^3 =0  \\ 
K_{\rm{sh}}p_{\rm{CH_4}}  p_{\rm{H_2O}} - p_{\rm{CO_2}}  p_{\rm{H_2}} =0 \end{cases}
 \end{equation}

\noindent
where
$K_{\mathrm{st}} = \exp\left(-\frac{E_{\mathrm{st}}}{RT}\right)$  \textendash ~equilibrium reaction constant equations for reforming reaction (Eq. (\ref{MSR})),
$E_{\rm{st}}$ \textendash ~energy activation for reforming reaction calculated as a function of temperature  [$\mathrm{\frac{J}{mol}}$],
$ K_{\mathrm{sh}} = \exp\left(-\frac{E_{\mathrm{sh}}}{RT}\right)$ \textendash ~equilibrium reaction constant equations for the shift reaction (Eq. (\ref{WGSR})),
$E_{\rm{sh}}$ \textendash ~energy activation for shift reaction calculated as a function of temperature [$\mathrm{\frac{J}{mol}}$]. The rate at which the mass of each chemical species is produced or consumed in both reactions was determined on the basis of reaction stoichiometry. This approach, consistent with the kinetic model, is illustrated in Table \ref{tab:mole_change}. Consequently, partial pressures for Eq. (\ref{eq:fgh}) can be calculated based on Table \ref{tab:mole_change} as follows:

\begin{equation}
\label{eq:p_ch4}
p_{\rm{CH_4}}=\frac{1-x}{1+SC+NC+2x}P
\end{equation}

\begin{equation} 
\label{eq:p_h2o}
p_{\rm{H_2O}}=\frac{SC-x-y}{1+SC+NC+2x}P
 \end{equation}

\begin{equation} 
\label{eq:p_h2}
p_{\rm{H_2}}=\frac{3x+y}{1+SC+NC+2x}P
 \end{equation}
 
 \begin{equation} 
\label{eq:p_co}
p_{\rm{CO}}=\frac{x-y}{1+SC+NC+2x}P
 \end{equation}

 \begin{equation} 
\label{eq:p_co2}
p_{\rm{CO_2}}=\frac{CC+y}{1+SC+NC+2x}P
 \end{equation}
where $P$ is total pressure [Pa], $SC$ is a ratio between supplied steam and methane, $NC$ is a ratio between supplied nitrogen and methane. 

\subsection{Data preparation}
\begin{figure}[h]
    \centering
    \includegraphics[width=32pc]{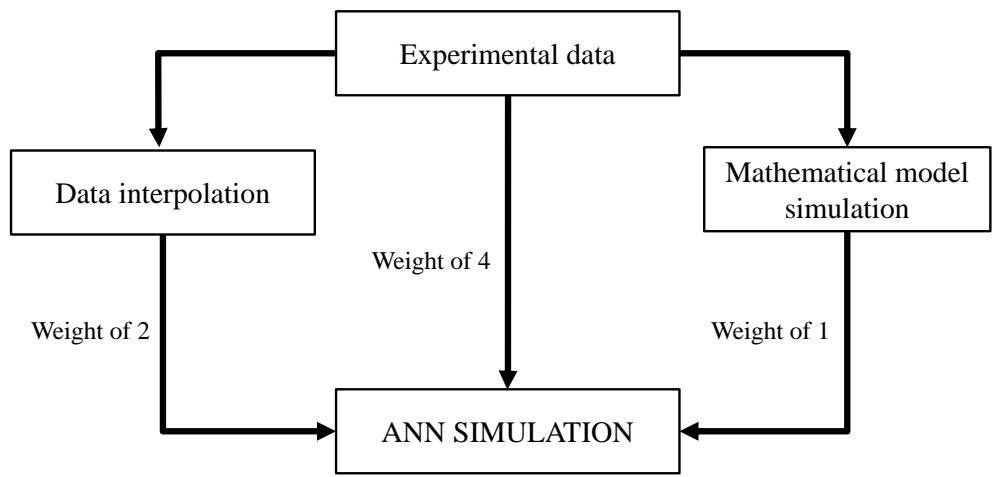}
    \caption{Presentation of used data augmentaton technique}
    \label{data_prep}
\end{figure}

The dataset was generated through a combination of experimental observations and mathematical modeling. It captures the composition of the post-reaction mixture as a function of key operating parameters, including process temperature, nickel catalyst mass, steam-to-methane ratio, nitrogen-to-methane ratio, and methane flow rate. To enhance the dataset's size and utility, the experimental data were interpolated using spline interpolation. Following this, a data augmentation technique was applied to emphasize the relative importance of different data types. Importance weights were assigned as follows: experimental data were given a weight of 4, interpolated data a weight of 2, and simulated data a weight of 1. This allocation reflects the higher reliability of experimental data compared to interpolated or simulated data, the latter being less precise due to simplifications inherent in the modeling process. Consequently, the dataset includes four times more records for experimental data than simulated data and twice as many interpolated data records as simulated data. The augmentation technique is illustrated in Figure \ref{data_prep}. Finally, the data were normalized to the range [0,1] to optimize the training process of the artificial neural network. The dataset utilized for training was randomly divided into three subsets: training data (70 \%), validation data (15 \%), and test data (15 \%). Additionally, an independent test dataset was employed exclusively to evaluate the performance of the ANN.

\subsection{ANN architecture selection}
The ANN was developed using MATLAB Software R2024b with the Deep Learning Toolbox.  The input layer of the feed-forward ANN consisted of five neurons, each corresponding to one of the operating parameters of the methane steam reforming process. The output layer contained four neurons, representing the four components of the post-reaction mixture, expressed in terms of dry gas composition. The log-sigmoid activation function was implemented in the hidden layers, except for the last layer. This activation function was selected due to its ability to map values to the range [0,1], facilitating the model's convergence during training. In the last layer, the softmax activation function was employed to produce a probability distribution, enhancing the model's physical interpretability. The Broyden-Fletcher-Goldfarb-Shanno Quasi-Newton Backpropagation algorithm was utilized for training the network, chosen for its efficiency in optimization and its ability to handle the complexity of the problem. Other hyperparameters were evaluated through the Random Sampling and the Bayesian optimization  to ensure the artificial neural network operated effectively. 

The evaluated hyperparameters included: number of hidden layers (ranging from one to four), neurons per layer (ranging from one to eight), learning rate (ranging from 0.0001 to 0.1) and number of epochs (ranging from 100 to 40,000). In the Bayesian Optimization, the expected-improvement-plus acquisition function was employed. The maximum allowable objective evaluations were set to 500. The optimization led to an ANN comprising three hidden layers with seven, eight, and eight neurons, respectively, a learning rate equal to 0.714, and a maximum epoch count of 16,261. A Mean Square Error (MSE) was a measure of training process performance and was equal to 0.000572. The network was also evaluated using an independent test dataset, with a Pearson and a Spearman correlation coefficients calculated to assess the alignment between predicted and actual values, resulting in 0.907 and 1.000, respectively. The results achieved through Bayesian Optimization were repeatable. The Random Sampling led to an ANN with an architecture consisting of three hidden layers with 6-8-6 neurons and was trained over 20,000 epochs with a learning rate equal to 0.001. The network was characterized by the following statistical parameter values: a MSE of 0.000498, a Pearson correlation coefficient of 0.927, and a Spearman correlation coefficient of 1.000. The obtained correlation coefficients were the highest, while the MSE was the lowest among all networks evaluated using both methods.

Moreover, predictions for unseen conditions were compared with the test data, and the results for the best-performing architecture are presented graphically in Section \ref{results}.  Furthermore, the selection of the optimal architecture accounted for the smoothness of the prediction functions, with particular attention to the continuity of their derivatives. Once again, network selected through Random Sampling demonstrated the best performance in this regard. This multi-faceted evaluation ensured that the chosen architecture provided robust and reliable predictions. 

\section{Results}
\label{results}
The results of the simulations performed using the chosen ANN architecture were compared with the test data to validate the precision of the model. Presented cases offer a comprehensive view of the model's performance and its ability to replicate real-world process dynamics.

\begin{figure}[H]
    \centering
    \includegraphics[width=1\linewidth]{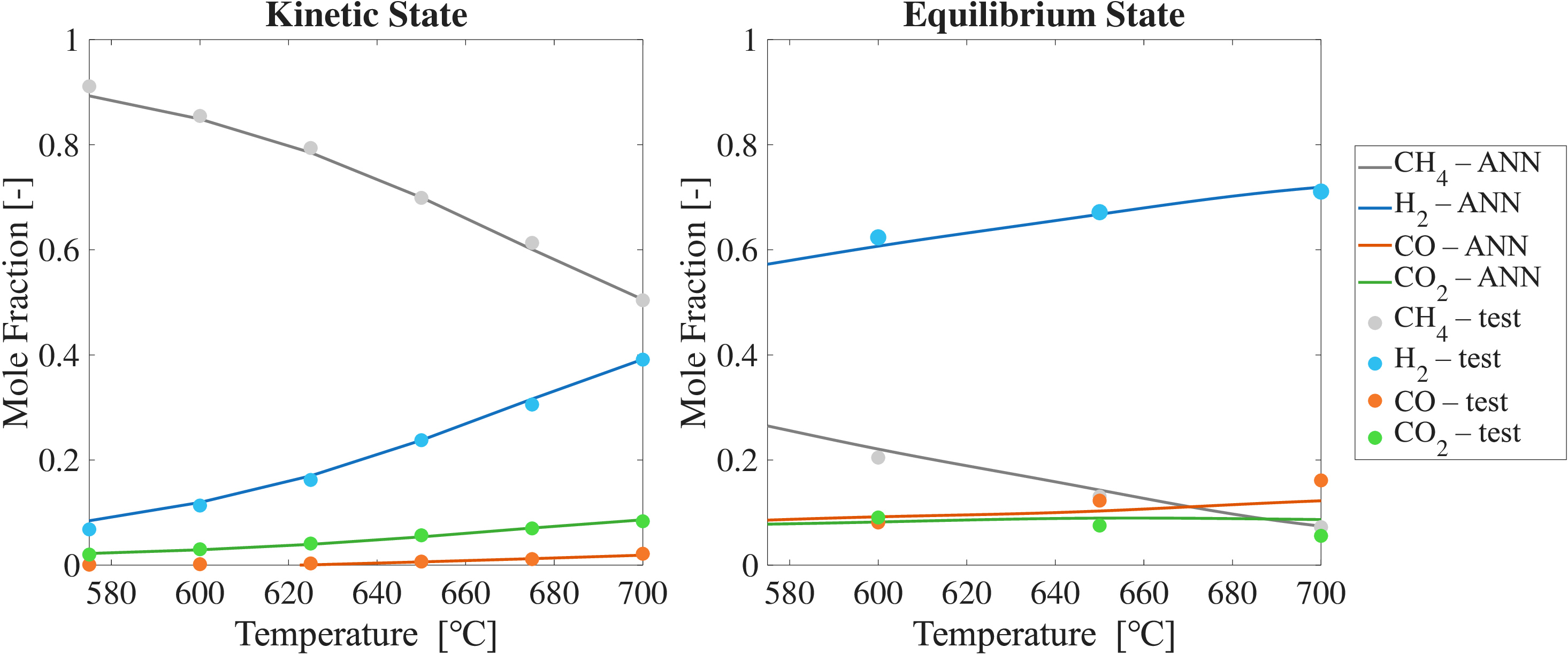}
    \caption{Comparison of ANN simulation results with test data (experimental data) for variable temperature values in kinetic state at $SC=3.00$, $NC=3.00$, $f_{\mathrm{CH_4}}=3.38\cdot 10^{-5}$ $\mathrm{\frac{mol}{s}}$, $m_{\mathrm{Ni}}=1.48$ $\mathrm{g}$ and in equilibrium state at  $SC=2.00$, $NC=0.00$, $f_{\mathrm{CH_4}}=1.01\cdot 10^{-4}$ $\mathrm{\frac{mol}{s}}$, $m_{\mathrm{Ni}}=5.03$ $\mathrm{g}$}
    \label{temperature}
\end{figure}

Figure \ref{temperature} illustrates a comparison between the ANN simulation results for post-reaction compound concentrations and experimental data across varying temperatures, with other parameters held constant, in both kinetic and equilibrium states. The experimental data are represented as points, while the ANN simulations are shown as continuous lines. A strong agreement is observed between the experimental and simulation results. In both states, increasing the temperature leads to an increase in H$_2$ and CO concentrations, while the concentration of CH$_4$ decreases.

\begin{figure}[H]
    \centering
    \includegraphics[width=1\linewidth]{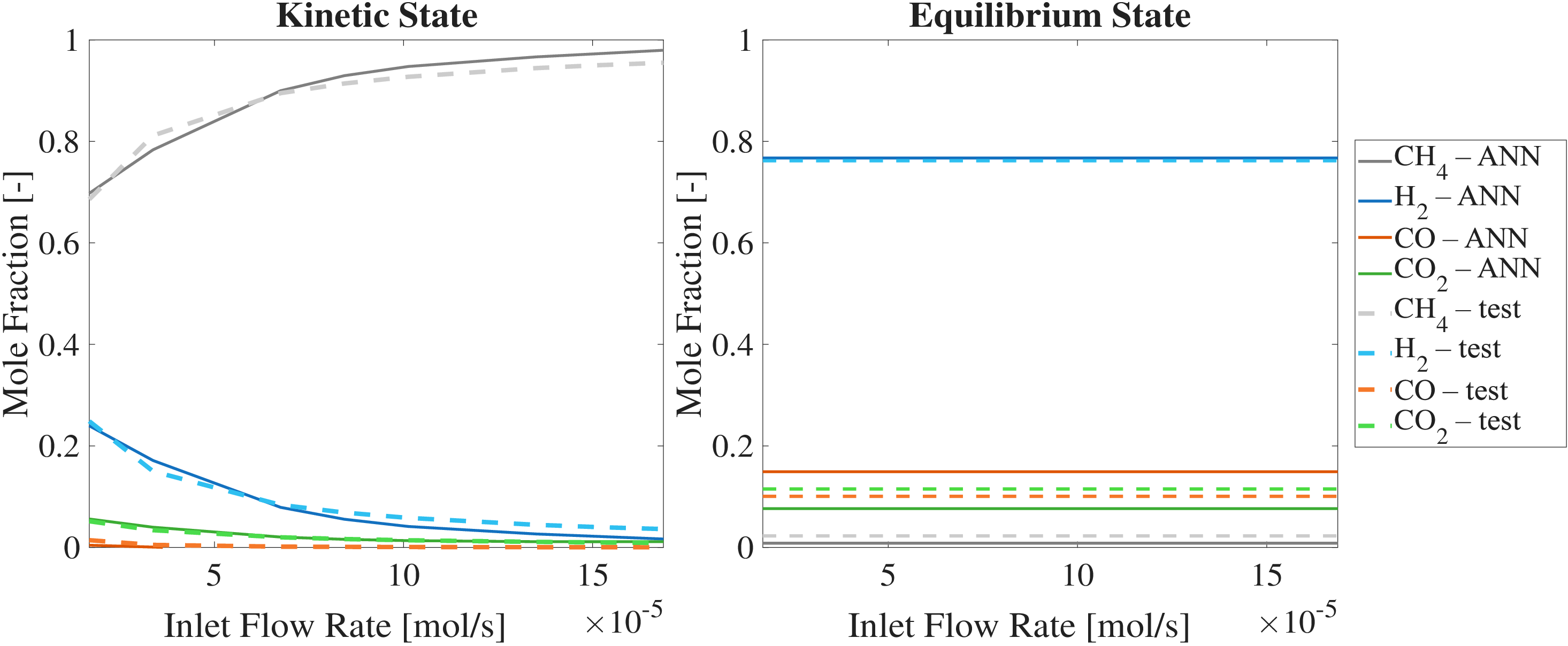}
    \caption{Comparison of ANN simulation outcomes with test data (from mathematical model simulations) for differing methane flow rates in kinetic state at $T=625$~$^\circ$C, $NC=3.00$, $SC=3.00$, $m_{\mathrm{Ni}}=1{.}50$ $\mathrm{g}$ and in equilibrium, state at $T=625$~$^\circ$C, $NC=3.00$, $SC=3.00$, $m_{\mathrm{Ni}}=15.00$ $\mathrm{g}$}
    \label{flow}
\end{figure}

Figure \ref{flow} presents a comparison of ANN simulation results with test data, generated using mathematical models, for post-reaction compound concentrations under varying inlet methane flow rates and constant other parameters in both kinetic and equilibrium states. The test data are shown as dashed lines, and the ANN simulations as continuous lines. The results demonstrate a good agreement between all datasets. In the kinetic state, an increase in methane flow rate leads to a decrease in hydrogen and carbon dioxide concentrations and an increase in methane concentration. In the equilibrium state, changes in the inlet flow rate do not significantly affect compound concentrations.

\begin{figure}[H]
    \centering
    \includegraphics[width=1\linewidth]{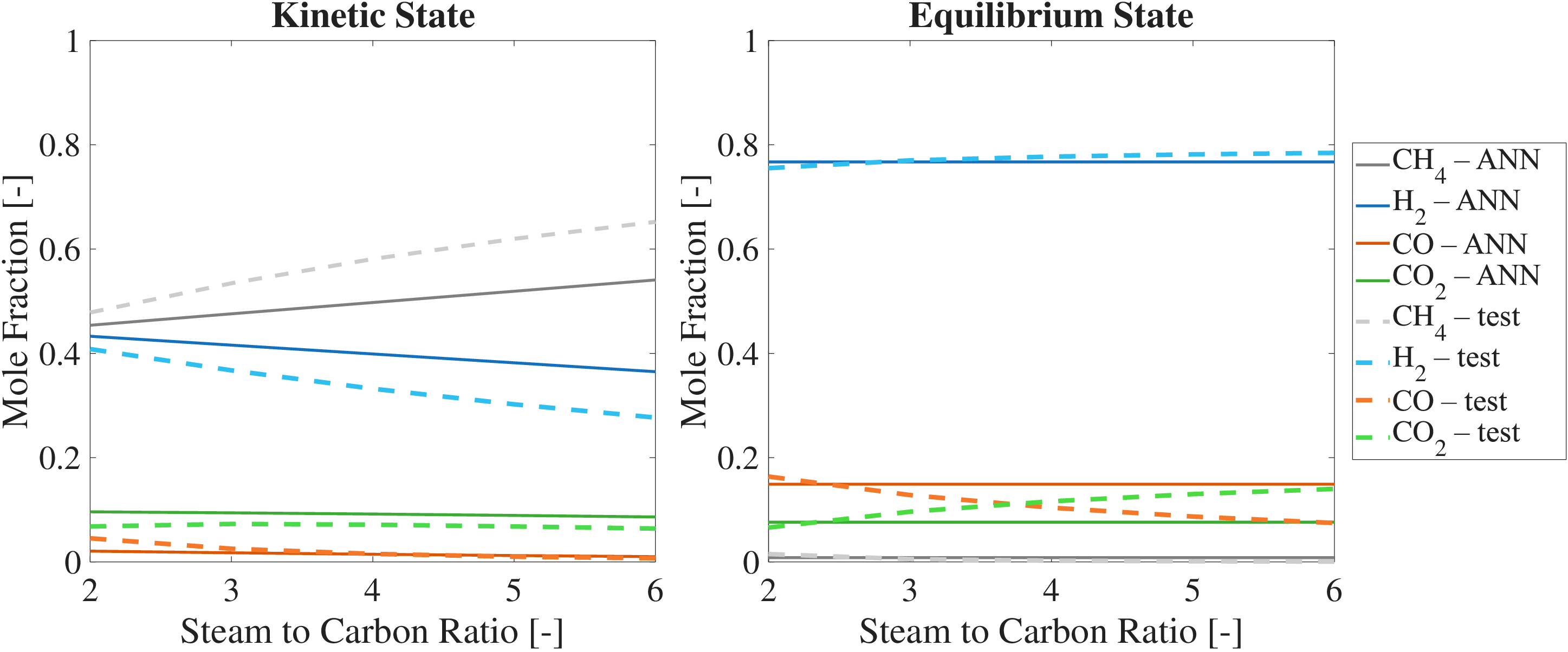}
    \caption{Comparison of ANN simulation outcomes with test data (from mathematical model simulations) for differing steam-to-carbon ratio in kinetic state at $T=700$~$^\circ$C, $NC=1.00$, $f_{\mathrm{CH_4}}=3.38\cdot 10^{-5}$ $\mathrm{\frac{mol}{s}}$, $m_{\mathrm{Ni}}=1{.}48$ $\mathrm{g}$ and in equilibrium, state at $T=700$~$^\circ$C, $NC=1.00$, $f_{\mathrm{CH_4}}=3.38\cdot 10^{-5}$ $\mathrm{\frac{mol}{s}}$, $m_{\mathrm{Ni}}=6.50$ $\mathrm{g}$}
    \label{SC}
\end{figure}

Figure \ref{SC} illustrates a comparison between ANN simulation results and test data derived from mathematical models. This comparison focuses on the concentrations of post-reaction compounds under a range of steam-to-carbon ratios, while keeping other parameters constant, for both the kinetic and equilibrium states. The test data are depicted as dashed lines, whereas the ANN simulations are shown as solid lines. The results indicate a strong alignment between the datasets. In the kinetic state, raising the steam-to-carbon ratio causes a reduction in hydrogen and carbon monoxide concentrations, with methane concentration rising. Conversely, in the equilibrium state, increasing the steam-to-carbon ratio leads to a decrease in methane and carbon monoxide concentrations, while hydrogen and carbon dioxide concentrations rise.

\begin{figure}[H]
    \centering
    \includegraphics[width=1\linewidth]{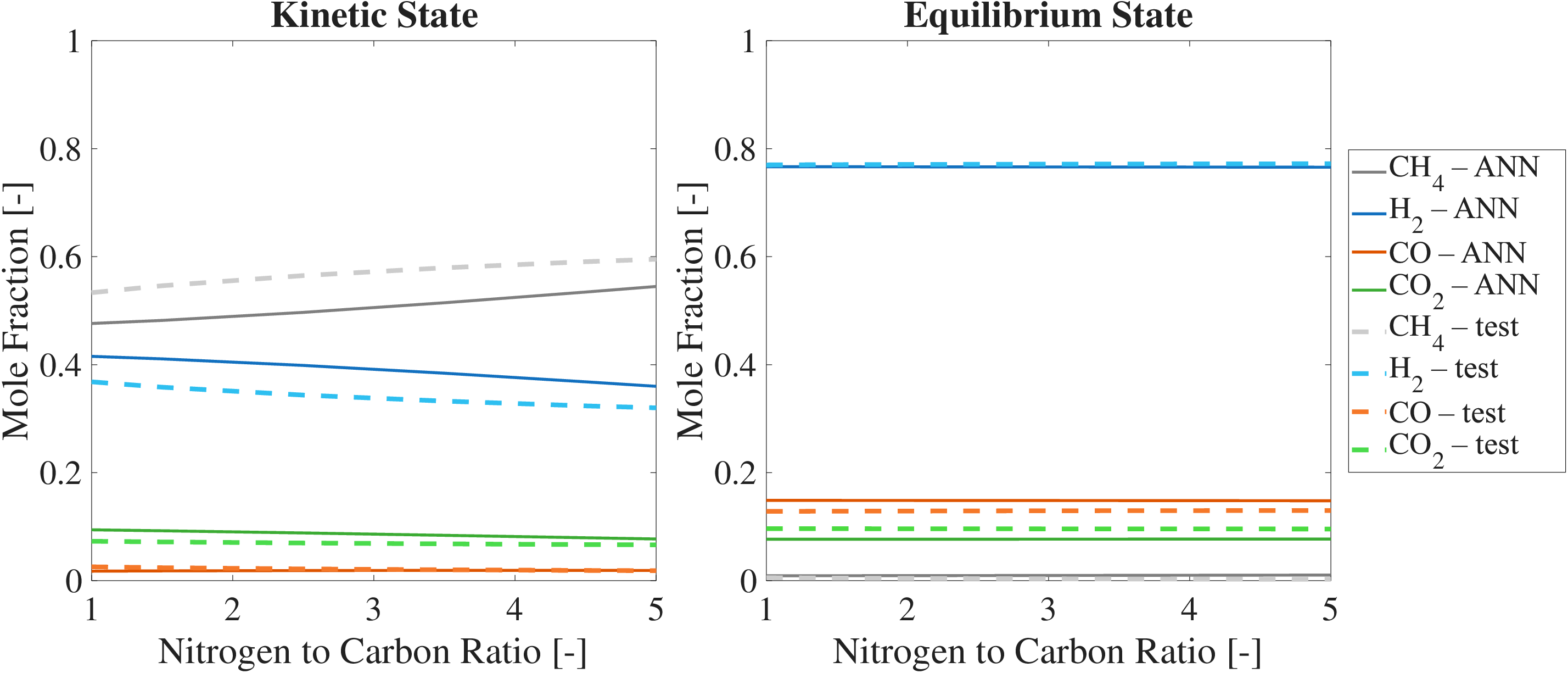}
    \caption{Comparison of ANN simulation outcomes with test data (from mathematical model simulations) for differing nitrogen-to-carbon ratio in kinetic state at $T=700$~$^\circ$C, $SC=3.00$, $f_{\mathrm{CH_4}}=3.38\cdot 10^{-5}$ $\mathrm{\frac{mol}{s}}$, $m_{\mathrm{Ni}}=1{.}48$ $\mathrm{g}$ and in equilibrium, state at $T=700$~$^\circ$C, $SC=3.00$, $f_{\mathrm{CH_4}}=3.38\cdot 10^{-5}$ $\mathrm{\frac{mol}{s}}$, $m_{\mathrm{Ni}}=6.50$ $\mathrm{g}$}
    \label{NC}
\end{figure}

Figure \ref{NC} depicts a comparison between ANN simulation outputs and test data obtained from mathematical models, focusing on the concentrations of reaction products across various nitrogen-to-carbon ratios, with other parameters held constant, for both kinetic and equilibrium states. In the figure, test data are represented by dashed lines, and ANN simulations are depicted by solid lines. The comparison reveals a strong correspondence between the datasets. In the kinetic state, an increase in the nitrogen-to-carbon ratio results in decreased concentrations of hydrogen, carbon monoxide, and carbon dioxide, while methane concentration increases. In contrast, under equilibrium conditions, altering the nitrogen-to-carbon ratio does not notably impact compound concentrations.

\begin{figure}[H]
    \centering
    \includegraphics[width=0.7\linewidth]{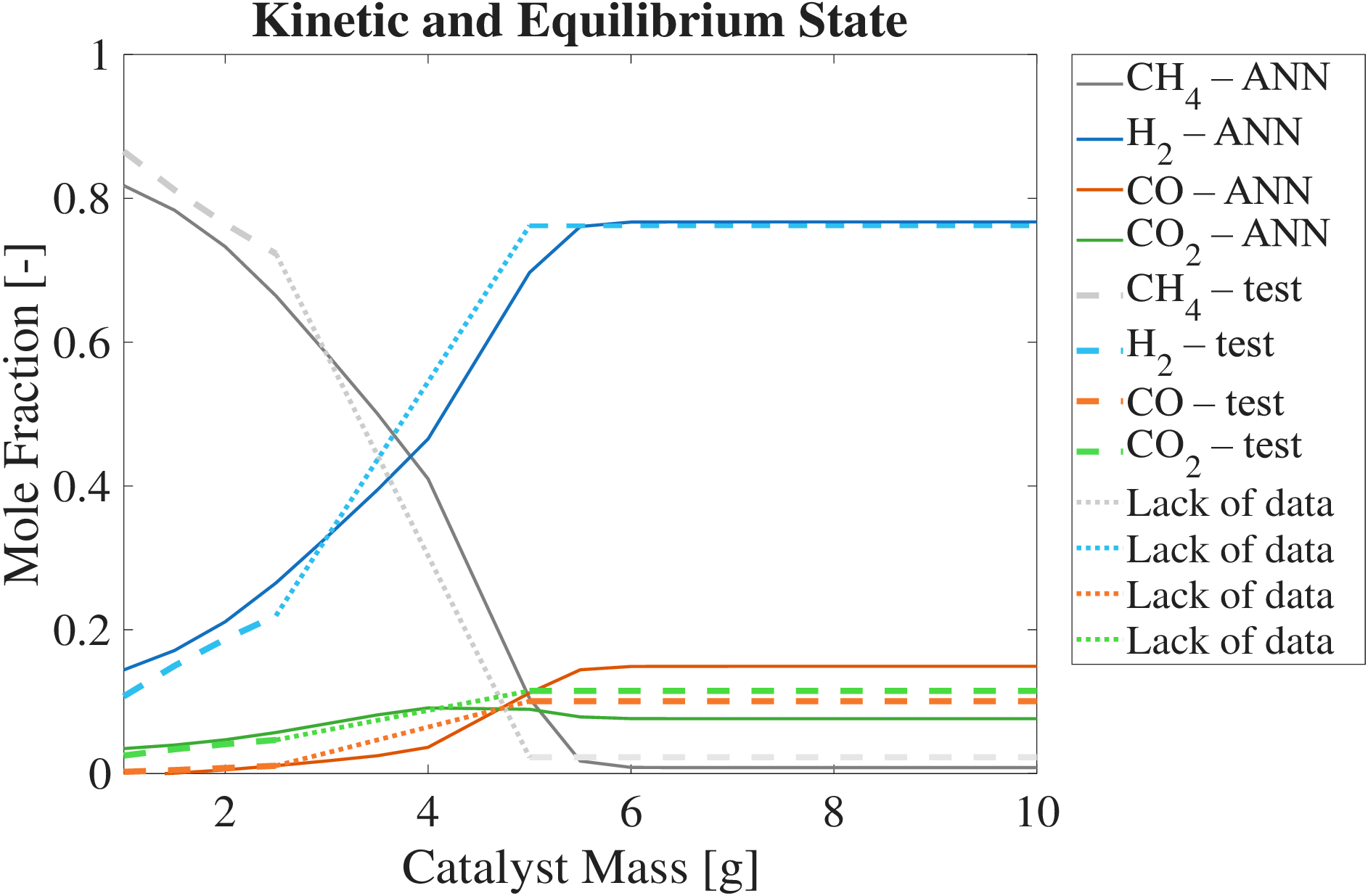}
    \caption{Comparison of ANN simulation results with test data (from mathematical model simulation) for variable catalyst mass values in kinetic and equilibrium state at $T=625~$$^\circ$C, $NC=3.00$, $SC=3.00$, $f_{\mathrm{CH_4}}=3.38\cdot 10^{-5}$ $\mathrm{\frac{mol}{s}}$}
    \label{nickel}
\end{figure}

Figure \ref{nickel} shows a comparison of ANN simulation results with test data, generated using mathematical models, for post-reaction compound concentrations under varying nickel catalyst masses while keeping other parameters constant, in both kinetic and equilibrium states. The test data are represented by dashed lines, while the ANN simulations are shown as continuous lines. A dotted line is included to indicate a lack of data, as this range of catalyst mass corresponds to a transition state for which no mathematical model was employed. The ANN predictions accurately represent the composition of the post-reaction system, including in the transition state. With increasing catalyst mass, the concentrations of H$_2$, CO, and CO$_2$ increase, while CH$_4$ decreases until equilibrium is reached, beyond which further changes in catalyst mass have no significant effect on concentrations.

\begin{figure}[H]
    \centering
    \includegraphics[width=0.6\linewidth]{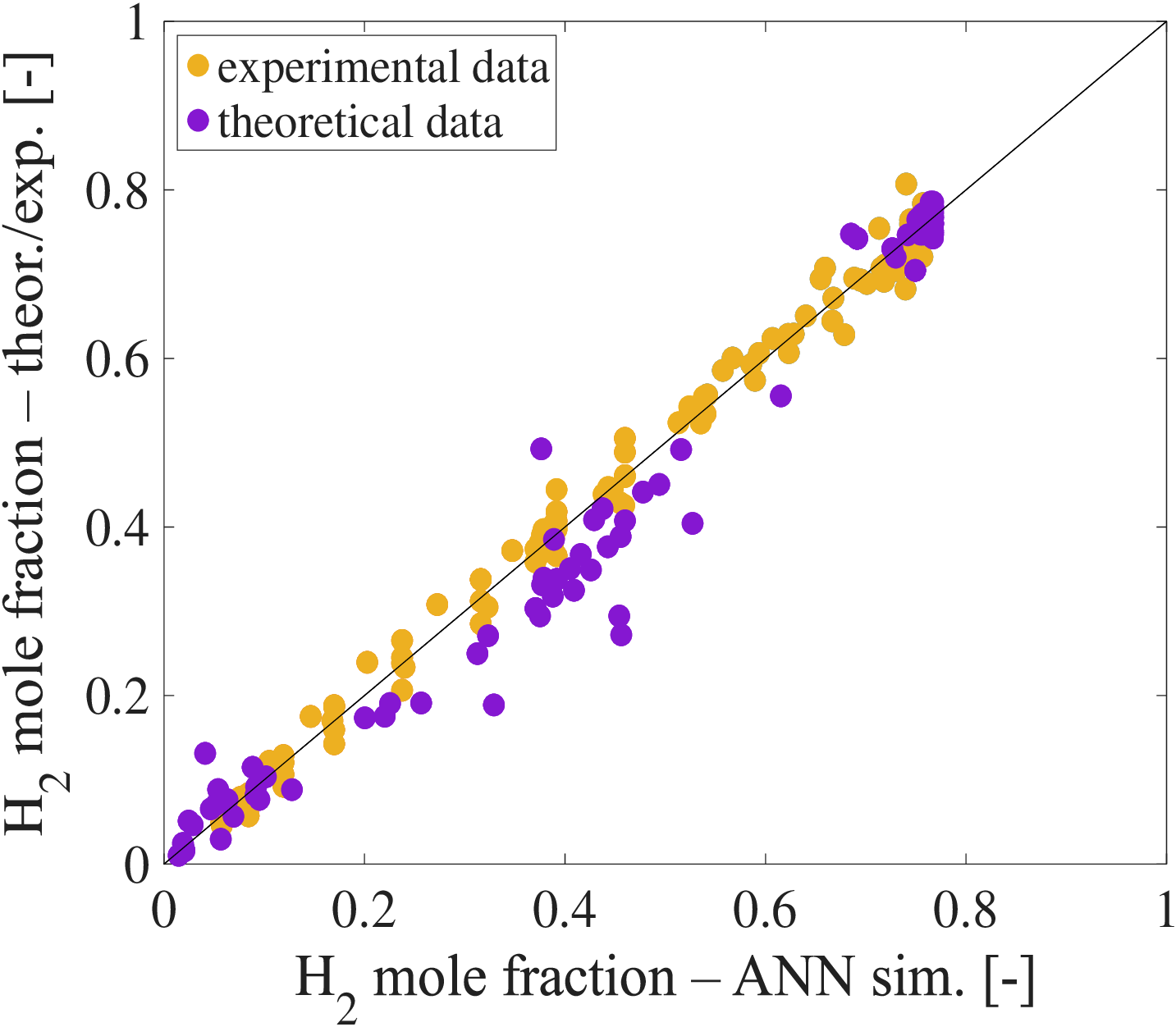}
    \caption{Comparison of experimental results, mathematical model simulations outcomes, and ANN simulation findings.}
    \label{exp_sim_ann}
\end{figure}

Figure \ref{exp_sim_ann} presents a comparison of hydrogen fractions in the post-reaction mixture as predicted by ANN simulations, theoretical mathematical models, and experimental measurements. The experimental data are represented as blue markers, while green markers correspond to the theoretical simulation results. The predictions made by the ANN closely match both experimental and theoretical data, with a stronger alignment with the experimental data, demonstrating the effectiveness of the network's training. This strong agreement demonstrates the reliability of the ANN in accurately predicting the hydrogen composition in the post-reaction mixture and highlights its potential as a robust tool for modeling complex chemical reaction systems.

\section{Conclusions} 
\label{conclusions}
An artificial neural network was developed to model the methane steam reforming process. The network was trained using a dataset comprising experimental, interpolated, and theoretical data, which was prepared using augmentation techniques. Bayesian Optimization and Random Sampling were employed to select the optimal ANN structure. The final network architecture consisted of an input layer with five neurons, three hidden layers with a 6-8-6 neuron configuration, and an output layer with four neurons. The learning rate was equal to 0.001. The training process was conducted over 20,000 epochs. The ANN demonstrated a high predictive accuracy for the composition of the post-reaction mixture under varying operating parameters, as evidenced by a mean squared error of 0.000498, a mean Pearson correlation coefficient of 0.927, and a Spearman correlation coefficient of 1.000. Furthermore, the ANN provided functions with continuous derivatives, enhancing its utility in process modeling and optimization. These results confirm that the designed ANN is a robust and reliable model for the methane steam reforming process under the investigated conditions.

\section*{Acknowledgments}
This research was partially funded by the Polish National Agency for Academic Exchange under the Strategic Partnerships Programme, Project No. BPI/PST/2021/1/00023 (project title: Strategic cooperation with Japan in the field of Energy and Environmental Engineering). The authors acknowledge support by Polish Ministry of Science and Higher Education program “Excellence Initiative—Research University” for the AGH University of Krakow, Poland and the Polish Ministry of Science and Higher Education Research Subsidy of the AGH University of Krakow for the Faculty of Energy and Fuels (Grant AGH no. 16.16.210.476). 

%Bibliography
\bibliographystyle{unsrt}  
\bibliography{bibliography}

\begin{thebibliography}{10}

\bibitem{iea}
{International Energy Agency}.
\newblock {Global Hydrogen Review 2024}, 2024.
\newblock Accessed: 24.11.2024.

\bibitem{rosen}
{Rosen MA, Koohi-Fayegh S}.
\newblock The prospects for hydrogen as an energy carrier: an overview of
  hydrogen energy and hydrogen energy systems.
\newblock {\em Energy Ecol. Environ.}, 1:10--29, 2013.

\bibitem{charles}
{Hawk CO, Golden PL, Storch HH, Fieldner AC}.
\newblock Conversion of methane to carbon monoxide and hydrogen.
\newblock {\em Ind. Eng. Chem. Res.}, 24:23--7, 1932.

\bibitem{HOLLADAY2009244}
{Holladay JD, Hu J, King DL, Wang Y}.
\newblock An overview of hydrogen production technologies.
\newblock {\em Catal. Today}, 139(4):244--60, 2009.
\newblock Hydrogen Production - Selected papers from the Hydrogen Production
  Symposium at the American Chemical Society 234th National Meeting \&
  Exposition, August 19-23, 2007, Boston, MA, USA.

\bibitem{en13040813}
{Zhu HL, Pastor-Pérez L, Millan M}.
\newblock Catalytic steam reforming of toluene: Understanding the influence of
  the main reaction parameters over a reference catalyst.
\newblock {\em Energies}, 13(4), 2020.

\bibitem{BAEK20149180}
{Baek SM, Kang JH, Lee KJ, Nam JH}.
\newblock A numerical study of the effectiveness factors of nickel catalyst
  pellets used in steam methane reforming for residential fuel cell
  applications.
\newblock {\em Int. J. Hydrog. Energy}, 39(17):9180--92, 2014.

\bibitem{BILGIC202318947}
{Bilgiç G, Bendeş E, Öztürk B, Atasever S}.
\newblock Recent advances in artificial neural network research for modeling
  hydrogen production processes.
\newblock {\em Int. J. Hydrog. Energy}, 48(50):18947--77, 2023.

\bibitem{ZAMANIYAN20136289}
{Zamaniyan A, Joda F, Behroozsarand A, Ebrahimi H}.
\newblock Application of artificial neural networks (ann) for modeling of
  industrial hydrogen plant.
\newblock {\em Int. J. Hydrog. Energy}, 38(15):6289--97, 2013.

\bibitem{NASR20133189}
{Nasr N, Hafez H, El Naggar MH, Nakhla G}.
\newblock Application of artificial neural networks for modeling of biohydrogen
  production.
\newblock {\em Int. J. Hydrog. Energy}, 38(8):3189--95, 2013.

\bibitem{GHASEMZADEH2018114}
{Ghasemzadeh K, Aghaeinejad-Meybodi A, Basile A}.
\newblock Hydrogen production as a green fuel in silica membrane reactor:
  Experimental analysis and artificial neural network modeling.
\newblock {\em Fuel}, 222:114--24, 2018.

\bibitem{GEORGE20189558}
{George J, Arun P, Muraleedharan C}.
\newblock Assessment of producer gas composition in air gasification of biomass
  using artificial neural network model.
\newblock {\em Int. J. Hydrog. Energy}, 43(20):9558--68, 2018.

\bibitem{HOSSAIN201611119}
{Hossain MA, Ayodele BV, Cheng CK, Khan MR}.
\newblock Artificial neural network modeling of hydrogen-rich syngas production
  from methane dry reforming over novel ni/cafe2o4 catalysts.
\newblock {\em Int. J. Hydrog. Energy}, 41(26):11119--30, 2016.

\bibitem{https://doi.org/10.1002/er.6483}
{Ayodele BV, Mustapa SI, Kanthasamy R, Zwawi M, Cheng CK}.
\newblock Modeling the prediction of hydrogen production by co-gasification of
  plastic and rubber wastes using machine learning algorithms.
\newblock {\em Int. J. Energy Res.}, 45(6):9580--94, 2021.

\bibitem{catal9090738}
{Ayodele BV, Mustapa SI, Alsaffar MA, Cheng CK}.
\newblock Artificial intelligence modelling approach for the prediction of
  co-rich hydrogen production rate from methane dry reforming.
\newblock {\em Catalysts}, 9(9), 2019.

\bibitem{AYODELE2021315}
{Ayodele BV, Alsaffar MA, Mustapa SI, Adesina A, Kanthasamy R, Witoon T,
  Abdullah S}.
\newblock Process intensification of hydrogen production by catalytic steam
  methane reforming: Performance analysis of multilayer perceptron-artificial
  neural networks and nonlinear response surface techniques.
\newblock {\em Process Saf. Environ. Prot.}, 156:315--29, 2021.

\bibitem{NKULIKIYINKA2020100037}
{Nkulikiyinka P, Yan Y, Güleç F, Manovic V, Clough PT}.
\newblock Prediction of sorption enhanced steam methane reforming products from
  machine learning based soft-sensor models.
\newblock {\em Energy and AI}, 2:100037, 2020.

\bibitem{VO2022820}
{Dat Vo N, Kang J-H, Oh D-H, Jung MY, Chung K, Lee C-H}.
\newblock Sensitivity analysis and artificial neural network-based optimization
  for low-carbon h2 production via a sorption-enhanced steam methane reforming
  (sesmr) process integrated with separation process.
\newblock {\em Int. J. Hydrog. Energy}, 47(2):820--47, 2022.

\bibitem{VO2019113809}
{Dat Vo N, Oh DH, Hong S-H, Oh M, Lee C-H}.
\newblock Combined approach using mathematical modelling and artificial neural
  network for chemical industries: Steam methane reformer.
\newblock {\em Appl. Energy}, 255:113809, 2019.

\bibitem{pizon}
{Pizoń Z, Kimijima S, Brus G}.
\newblock Enhancing a deep learning model for the steam reforming process using
  data augmentation techniques.
\newblock {\em Energies}, 17(10), 2024.

\bibitem{Brus.equil}
{Brus G, Nowak R, Szmyd JS, Komatsu Y, Kimijima S}.
\newblock An experimental and theoretical approach for the carbon deposition
  problem during steam reforming of model biogas.
\newblock {\em J. of theoretical and appl. mechanics}, 53(2):273--84, 2012.

\bibitem{brus.2012}
{Brus G}.
\newblock Experimental and numerical studies on chemically reacting gas flow in
  the porous structure of a solid oxide fuel cells internal fuel reformer.
\newblock {\em Int. J. Hydrog. Energy}, 37(22):17225--34, 2012.
\newblock HySafe 1.

\bibitem{Brus.2012b}
Grzegorz Brus, Y~Komatsu, Shinji Kimijima, and Janusz~S Szmyd.
\newblock {An analysis of biogas reforming process on Ni/YSZ and Ni/SDC
  catalysts}.
\newblock {\em International Journal of Thermodynamics}, 15(1):43 -- 51, 2012.
\newblock Export Date: 7 June 2019.

\bibitem{AGC2009Catalyst}
{AGC Semi Chemical Co.}
\newblock Catalyst analysis sheets.
\newblock AGC Semi Chemical Co., 2009.
\newblock Technical Data Sheets.

\end{thebibliography}

\end{document}